\documentclass[sigconf]{acmart}
\usepackage{enumitem}
\usepackage{multirow}

\AtBeginDocument{%
  \providecommand\BibTeX{{%
    \normalfont B\kern-0.5em{\scshape i\kern-0.25em b}\kern-0.8em\TeX}}}

\copyrightyear{2023}
\acmYear{2023}
\setcopyright{rightsretained}
\acmConference[CHI '23]{Proceedings of the 2023 CHI Conference on Human Factors in Computing Systems}{April 23--28, 2023}{Hamburg, Germany}
\acmBooktitle{Proceedings of the 2023 CHI Conference on Human Factors in Computing Systems (CHI '23), April 23--28, 2023, Hamburg, Germany}\acmDOI{10.1145/3544548.3581152}
\acmISBN{978-1-4503-9421-5/23/04}

\begin{document}

\title{A Mixed-Methods Approach to Understanding User Trust after Voice Assistant Failures}
\raggedbottom

\author{Amanda Baughan}
\authornote{This work was conducted as part of an internship with Google Research.}
\email{baughan@cs.washington.edu}
\affiliation{%
  \institution{University of Washington}
  \city{Seattle}
  \state{WA}
  \country{USA}
}

\author{Allison Mercurio}
\email{amercurio@google.com}
\affiliation{%
  \institution{Google Research}
  \city{Mountain View}
  \state{CA}
  \country{USA}
}

\author{Ariel Liu}
\email{arielliu@google.com}
\affiliation{%
  \institution{Google Research}
  \city{Mountain View}
  \state{CA}
  \country{USA}
} 

\author{Xuezhi Wang}
\email{xuezhiw@google.com}
\affiliation{%
  \institution{Google Research}
  \city{New York}
  \state{NY}
  \country{USA}
}

\author{Jilin Chen}
\email{jilinc@google.com}
\affiliation{%
  \institution{Google Research}
  \city{Mountain View}
  \state{CA}
  \country{USA}
}

\author{Xiao Ma}
\email{xmaa@google.com}
\affiliation{%
  \institution{Google Research}
  \city{New York}
  \state{NY}
  \country{USA}
}

\renewcommand{\shortauthors}{Baughan et al.}



\begin{abstract}
  Despite huge gains in performance in natural language understanding via large language models in recent years, voice assistants still often fail to meet user expectations. In this study, we conducted a mixed-methods analysis of how voice assistant failures affect users' trust in their voice assistants. To illustrate how users have experienced these failures, we contribute a crowdsourced dataset of 199 voice assistant failures, categorized across 12 failure sources. Relying on interview and survey data, we find that certain failures, such as those due to overcapturing users' input, derail user trust more than others. We additionally examine how failures impact users' willingness to rely on voice assistants for future tasks. Users often stop using their voice assistants for specific tasks that result in failures for a short period of time before resuming similar usage. We demonstrate the importance of low stakes tasks, such as playing music, towards building trust after failures.

\end{abstract}

\maketitle

\section{Introduction}
Voice assistants have received a lot of attention from both industry and academia, especially given the recent advances in natural language processing (NLP). Within the past five years, advancements in NLP have achieved huge gains in accuracy when tested against standard datasets~\cite{devlin2018bert, liu2019roberta, brown2020language, vaswani2017attention, smith2022using, wei2021finetuned, thoppilan2022lamda}, with state-of-the-art accuracy in natural language processing models as high as 99\% for certain tasks~\cite{liu2019roberta, brown2020language}.
This has led many practitioners and researchers alike to imagine a near future where voice assistants can be used in increasingly complex ways, including supporting healthcare tasks~\cite{sezgin2020readiness, mehandru2022reliable}, giving mental health advice~\cite{yang2021clinical, saha2022shoulder}, and high stakes decision-making~\cite{de2020reducing}.

However, despite the increasing accuracy of NLP models and the breadth of their applications, evidences suggest that users remain reluctant and distrusting of using voice assistants~\cite{voiceassistantsnotused2017, Sirialexanotused2022}.
In the U.S., voice assistants are common in homes, with an estimated 72\% of Americans having used a voice assistant~\cite{pwc_voice}.
However, people primarily use these for basic tasks such as playing music, setting timers, and making shopping lists~\cite{voiceassistantsnotused2017, Sirialexanotused2022, luger2016like}. This is because when voice assistants fail, such as by incorrectly answering a question, it derails user trust~\cite{pwc_voice, luger2016like}. User trust is pivotal to user adoption of various technologies~\cite{bahmanziari2003trust}, and in this case, low user trust results in reluctance to try voice assistants' novel capabilities.

As voice assistants increasingly rely on large language models ~\cite{googleblog2021, FitzGerald2022}, we believe the gap between the high accuracy of these models and users' reluctance to use voice assistants for complex tasks may be explained by differences in how users and NLP practitioners evaluate the success of a model.
Standard NLP models are often evaluated on large datasets of coherent text-based questions and answers~\cite{rajpurkar2016squad, rajpurkar2018know} or paired written dialogue~\cite{zhang2018personalizing}. Meanwhile, in practice users' speech may include disfluencies, such as restarts and filler words, questions not covered in training, or background noise which misconstrues speech. 
In the case of question answering, NLP models are evaluated based on how many questions are accurately answered on a subset of the training dataset~\cite{rajpurkar2016squad, rajpurkar2018know}.
As one may expect, people can interact with voice assistants in a multitude of ways that fall outside of the scope of training data, which can lead to friction.  In the eyes of users, these inaccurate responses, or voice assistant failures, can lead to frustration. For example, only five percent of users report never becoming frustrated when using voice search~\cite{cox_2020_voicesearch}.

We believe that the gap between how NLP models are evaluated and how users encounter and perceive failures hinders the practical applications of the advancements that voice assistants have made.
Therefore, we ask, which types of voice assistant failures do users currently experience, and how do these failures affect user trust?
A human-centered understanding of the types of NLP failures that occur and their impact on users trust would allow technologists to prioritize and address critical failures and enable long-term adoption of voice assistants for a wider variety of use cases.

Further, while research has started to categorize types of breakdowns in communication between users and NLP agents~\cite{Hong2021nlpfailures, paek2013conversation}, little work has looked into how users perceive these failures and subsequently trust and use their voice assistants.
We draw from and extend past research to make the following contributions:
\begin{itemize}[leftmargin=.3in]
    \item \textbf{C1:} Iterating on the existing taxonomy of NLP failures, we crowdsource a dataset of 199 failures users have experienced across 12 different sources of failure.
    \item \textbf{C2:} A qualitative and quantitative evaluation on how these different failures affect user trust, specifically along dimensions of ability, benevolence, and integrity.
    \item \textbf{C3:} A qualitative and quantitative analysis on how trust impacts intended future use.
\end{itemize}

To accomplish this, we developed a mixed-methods, human-centered investigation into voice assistant failures.
We first executed interviews with 12 voice assistant users to understand what types of failures they have experienced and how this affected their trust and subsequent use of their assistant. 
We concurrently crowdsourced a dataset of failures from voice assistant users on Amazon Mechanical Turk.
Finally, we executed a survey to quantify how different types of failures impact users' trust in their voice assistants and their willingness to use them for various tasks in the future. 

We found that different types of voice assistant failures have a differential impact on trust. Our interviews and survey revealed that participants are more forgiving of failures due to spurious triggers or ambiguity of their own request. In the case of spurious triggers, the voice assistant activates due to mishearing the activation phrase when it was not said. Users forgave this more easily, as it did not hinder them from accomplishing a goal. Failures due to ambiguity occurred when there were multiple reasonable interpretations of a request, and the response was misaligned with what the user intended while still accurately answering the question. Users tended to blame themselves for these failures. However, failures due to overcapture more severely reduced users' trust, as when the voice assistant continued listening without any additional input, users considered their use a waste of time. 

We additionally find that on many occasions, users would discontinue using their voice assistant for a specific task for a short period of time following a failure, and then resume again once trust had been rebuilt. Trust was often rebuilt by using the voice assistant for tasks they considered simple, such as playing music, or alternatively, using the voice assistant for the same general task but in a different use case. 
In addition to these findings, we release a dataset of 199 voice assistant failures, capturing user input, voice assistant response, and the context for the failure, so that researchers may use these failures for future research on how users respond to voice assistant failures.
As voice assistants continue to perform increasingly complex and high stakes tasks across various industries~\cite{mehandru2022reliable, sezgin2020readiness, de2020reducing, yang2021clinical, robertson2022understanding}, we hope that this research will help technologists understand, prioritize, and address natural language failures to increase and maintain user trust in voice assistants.

\section{Related Work}

Prior research across many fields has examined the interaction between users and voice assistants, including human-computer interaction, human-centered AI, human-robotics interaction, science and technology studies (STS), computer-mediated communication (CMC), and social psychology. 
In addition, some work in natural language processing (NLP), especially NLP robustness, has approached technology failures in voice assistants and developed certain technical solutions to address them.
Here, we provide an interdisciplinary review of research relevant to voice assistant failures during user interaction across these fields.
The literature review is organized as follows:
1) literature on user expectations and trust in voice assistants;
2) human-computer interaction (HCI) approaches to understanding voice assistant failures and strategies for mitigation;
3) natural language processing (NLP) approaches to voice assistant failures, including disfluency and robustness.

\subsection{User Expectations and Trust in Voice Assistants}

Researchers have long tried to understand how people interact with automated agents, especially comparing and contrasting these experiences with human-to-human communication.
When talking with other humans, conversations can broadly be understood as functional (also known as transactional or task-based) or social (interactional), and many conversations include a mix of both~\cite{clark2019goodconvo}. Functional conversations serve towards the pursuit of a goal, and those who participate often have understood roles towards the pursuit of that goal. In contrast, social conversations have a goal of building, strengthening, or maintaining a positive relationship with one of the participants. These social conversations can help build trust, rapport, and common ground~\cite{clark2019goodconvo}. 

People generally expect to have functional conversations with voice assistants~\cite{clark2019goodconvo}. The lack of social conversations may reduce users' ability to build trust in their voice assistants.
Indeed, past research has shown that users trust embodied conversational agents more when they engage in small talk~\cite{bickmore2001relational}, although this varies by user personality type and level of embodiment of the agent~\cite{bickmore2005social}.
As it stands, people report not using voice assistants for a broad range of tasks, even though they're technically capable of doing so~\cite{pwc_voice}. 
Prior work has illustrated the importance of trust for continued voice assistant use~\cite{luger2016like, lahoual2019users}, as trust is pivotal to user adoption of voice assistants~\cite{nasirian2017ai, lee2021role} and willingness to broaden the scope of voice assistant tasks~\cite{pwc_voice}.
It is especially important to support trust-building between users and voice assistants as researchers continue to imagine and develop new capabilities for them, including complex tasks such as supporting healthcare tasks~\cite{sezgin2020readiness, mehandru2022reliable}, giving mental health advice~\cite{yang2021clinical, saha2022shoulder}, and other high stakes decision-making~\cite{de2020reducing}.

This then begs the question of how trust is built between users and voice assistants.
Trust in machines is an increasingly important topic, as use of automated systems is widespread~\cite{winter2022if}.
Concretely, trust can be conceptualized as a combination of confidence in a system as well as willingness to act on its provided recommendations~\cite{seymour2021exploring, Madsen00measuringtrust}. Prior researchers have examined trust in machines in terms of people's confidence in a machine's \textit{ability} to perform as expected, \textit{benevolence} (well-meaning), and \textit{integrity} to adhere to ethical standards~\cite{ma2017self}
Broadly, past research has evaluated how various factors such as accuracy and errors affect people's trust in algorithms~\cite{yin2019understanding,dzindolet2002perceived, dietvorst2015algorithm}.
In the case of voice assistants, \citet{nasirian2017ai} and \citet{lee2021role} studied how quality affects trust in and adoption of voice assistants, and found that information and system quality did not impact users' trust in a voice assistant, but interaction quality did. Interaction quality was captured based on a study by \citet{ekinci2009consumer}, in which Likert scale responses were captured regarding competence, attitude, service manner, and responsiveness of the voice assistant. 
In addition, customizing a voice assistant's personality to the user can lead to higher trust~\cite{braun2019your}, while gender does not impact users' trust in a voice assistant~\cite{tolmeijer2021female}. Overall, prior research demonstrates the importance of the interaction quality and social conversations for building trust between users and voice assistants, which in turn affects users' willingness to continue using them and broaden the scope of their tasks.

\subsection{HCI Approaches to Voice Assistant Failures}
However, there are occasionally unforeseen breaches of trust, as not all interactions go as smoothly as one expects.
Prior work has explored the diversity of issues affecting engagement and ongoing use of voice assistants and has shown that when users have expectations for voice assistants that surpass its capabilities, voice assistant failures and user frustration ensues~\cite{luger2016like, lahoual2019users}.

This begs the question, how has prior work defined failures in voice assistants? Some work uses specific scenarios in their studies. For example, \citet{lahoual2019users} conducted evaluation in domestic and driving situations. They identified failures due to poor voice recognition, limited understanding of a command, and connectivity.
\citet{cuadra2021my} used failures in specific tasks, such as attempting to give directions to an incorrect location, send a text to the wrong person, play the wrong type of music, or adding a reminder with an incorrect detail~\cite{cuadra2021my}.
\citet{mahmood2022owning} simulated online shopping, in which an AI assistant with a voice component would fail by using homonyms of the requested items. For example, the ambiguous item ``bow'' could mean a hair bow, archery bow, or bow for gift wrapping. \citet{salem2015would} had participants control a robot's movement, and in the faulty condition, the robot would move erratically, incorrectly responding to the users' input. \citet{candello2019audiences} defined failure as occasions in which someone asked a question that could not be understood or was out of scope of the voice assistants' knowledge, in which case it would divert the conversation to ask an unrelated question.

Other research aims to provide a broad categorization of voice assistant failures, drawing from
theoretical frameworks of communication between humans~\cite{Hong2021nlpfailures, paek2013conversation, clark1996using}.
We reference Herbert Clark's grounding model for human communication, which relies on four different levels to achieve mutual understanding: channel, signal, intention, and conversation~\cite{clark1996using}. 
This was expanded by \citet{paek2013conversation}, which applied these four levels to human-machine interactions and failure points. Channel level errors include when an AI fails to attend to a users' attempt to initiate communication; signal level errors include an error in capturing user input (e.g. due to transcription); intention level errors include mistakes in making sense of the semantic meaning of the transcribed input; and conversation level errors occur when a user has requested an unknown action to the AI (e.g. asking a weather app to schedule something).
\citet{Hong2021nlpfailures} built on this model, specifically restricting the context to NLP failures, rather than AI as a whole. Based on interviews with NLP practitioners, they renamed the categories as attention (channel), perception (signal), understanding (intention), and response (conversation). \citet{Hong2021nlpfailures} focused on failures that are either very common, or rare but very costly, to cover the most important and frequent failures users encounter when interacting with NLP-based systems. In this work, we build on their existing taxonomy of NLP failures~\cite{Hong2021nlpfailures}, narrowing the use case to only voice assistant failures, and evaluating how different failures impact on user trust and future intended use.

There is currently little systematic evaluation of the impact of voice assistant failures on user trust.
\citet{salem2015would} found that if a robot had faulty performance, this did not influence participants’ decisions to comply with its requests, but it did significantly affect their perceptions of the robot's reliability and trustworthiness. 
\citet{mahmood2022owning} found that voice assistants that accepted blame and apologized for mistakes were thought to be more intelligent, likeable, and effective in recovering from failures than assistants that shifted the blame.

Sometimes after a failure, users will try to reformulate, simplify, or hyper-enunciate their commands as a way to continue using the device~\cite{lahoual2019users, luger2016like, myers2018patterns, velkovska2018illusion}.
If users are repeatedly unable to repair failures with voice assistant, this weakens their trust and causes them to reduce their scope of commands to simple tasks with low risk of failure~\cite{luger2016like, lahoual2019users}. 
~\citet{lahoual2019users} found that in some situations, voice assistant failures can erode trust to the extent that users abandon voice assistants all together.
However, not all failures require self-repair.
A study by \citet{cuadra2021my} found that when voice assistants make mistakes, voice assistant self-repair greatly improves people’s assessment of an intelligent voice assistant, but it can have the opposite impact if no correction is needed.
Thus, understanding which types of failures undermine trust the most may also inform us when failure mitigation strategies should be activated.

\begin{figure*}[htbp]
    \centering
    \includegraphics[width=.75\textwidth]{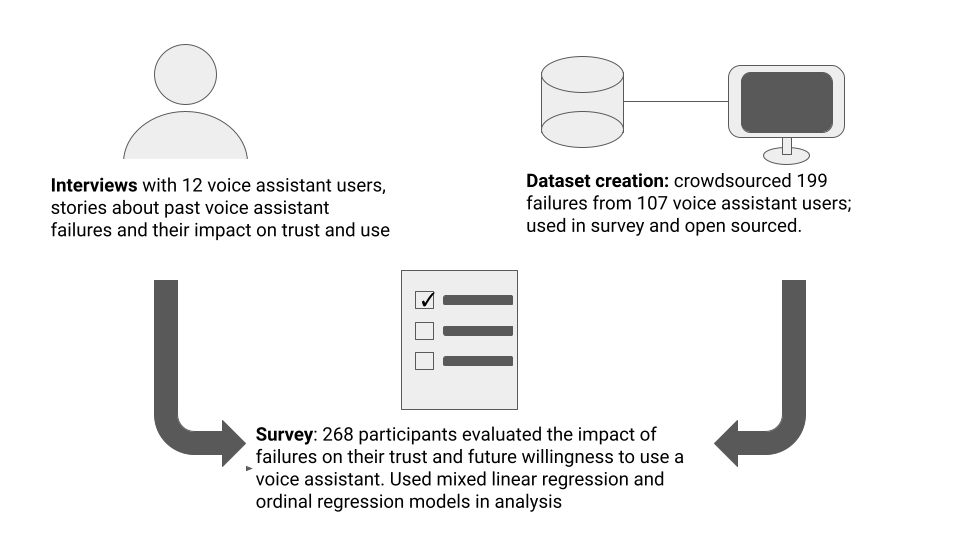}
    \caption{To analyze the impact of voice assistant failures on user trust, we used a mixed-methods approach, including interviews and a survey. As part of the materials for our survey, we crowdsourced 199 failures from 107 voice assistant users, and include this dataset as part of our contributions.}
    \label{fig:my_label}
    \Description{This is a visual aid to show that our paper contained three core elements: the interviews, dataset creation, and survey, which are each represented with a visual and text on the figure. On the top left, interviews with 12 voice assistant users are represented with a geometric person, stating "Stories about past voice assistant failures and their impact on trust and use." This has an arrow pointing towards the survey. On the top right, the dataset creation is represented with a cylinder and a computer, stating "crowdsourced 199 failures from 107 voice assistant users, used in survey and open-sourced." This also has an arrow leading to the survey at the bottom. The survey is represented by a sheet of paper with a checklist and states ``268 participants evaluated the impact of failures on their trust and future willingness to use a voice assistant. Used mixed linear regression and ordinal regression models in analysis.''}
\end{figure*}

\subsection{NLP Approaches to Voice Assistant Failures}
The NLP community has also examined voice assistant failures from a slightly different angle, focusing on the robustness of different NLP components underlying voice assistants, such as models for tasks in natural language inference~\cite{naik2018stress}, question answering~\cite{gupta-etal-2021-disflqa, miller2020effect}, and speech recognition~\cite{lee2018spoken}.
NLP robustness can be defined as understanding how model performance changes when testing on a new dataset, which has a different distribution from the dataset the model is trained on~\cite{wang2021measure}.
In practice, users' real world interactions with voice assistants could differ from data used in development, which mimics the data distribution shift in NLP robustness research.

Such data distribution shifts are shown to lead to model failures.
In the case of question answering, state-of-art models perform nearly at human-level for reading comprehension on standard benchmarks collected from Wikipedia~\cite{rajpurkar2016squad}.
However, 
\citet{miller2020effect} found that model performance drops when the question answering model is evaluated on different topic domains, such as New York Times articles, Reddit posts, and Amazon product reviews.
Noisy input can also harm model performances.
\citet{lee2018spoken} showed speech recognition errors have catastrophic impact on machine comprehension.
\citet{gupta-etal-2021-disflqa} created a question answering dataset Disflu-QA where humans introduce contextual disfluencies, which also lead to model performance drops.

Although these works do not directly focus on voice assistant failures, topic domain changes, speech recognition errors and disfluencies are all very common during user interactions with voice assistants.
Such similarities motivate us to draw parallels between the NLP robustness literature and HCI perspectives of system failures.
By understanding how different types of failures affect trust in voice assistants overall, we can then try to pinpoint the underlying NLP components that are the root cause of the most critical failures that erode trust~\cite{khaziev2022fpi}.
Technical solutions can then be leveraged to improve the robustness of the most critical parts of the system in order to increase user trust and long-term engagement most efficiently.

\section{Method Overview}
Now that we have established the importance of understanding of how voice assistant failures impact user trust, we proceed to conduct a mixed-method study.
First, to prepare for the quantitative evaluation, we reviewed existing datasets in HCI and NLP to find failures that we could use as materials for our survey. Ultimately, the existing datasets were not sufficient for our needs.
Therefore, we crowdsourced a dataset of failures from voice assistant users, which we also open source as part of the contributions of this study.
Concurrently, we conducted interviews with 12 voice assistant users to understand which types of failures they have experienced, and how this affected their trust in and subsequent use of the assistant.
These interviews were designed to provide a broad understanding of the thoughts, feelings, and behaviors that users have with regard to voice assistant failures and inform the quantitative survey design.
Finally, we executed a survey to quantify how different types of failures impact user perceptions of trust in their voice assistants and their willingness to use them for various tasks in the future. To report these findings, we first describe our process of collecting the crowdsourced dataset of failures, and how we selected a subset to use in our survey. Next, we present the interviews and survey, first describing our data collection and analysis, and then presenting the results concurrently.

\section{Crowdsourcing a Dataset of Voice Assistant Failures}

The first goal in our investigation was to determine which types of failures users experience when using voice assistants. We first evaluated existing datasets for fit and breadth of failures. We determined they were not sufficient for our purposes, so we proceeded to crowdsource a dataset of failures, adapting a taxonomy from \citet{Hong2021nlpfailures} to guide our collection. Finally, we cleaned and open-sourced this dataset as a contribution of our work.

\begin{table*}[htbp]
    \centering
    \small
    \begin{tabular}{p{0.12\linewidth} | p{4cm} | p{0.15\linewidth} | p{0.35\linewidth}} 
      \textbf{Failure Type} & \textbf{Sequential Coding Guide} & \textbf{Failure Source} & \textbf{Failure Scenario} \\ 
      \hline
      \hline
    \multirow{7}{*}{\begin{tabular}[c]{@{}l@{}}\textbf{Attention}\end{tabular}} & \multirow{7}{*}{\begin{tabular}[c]{@{}l@{}}A lack of visual or audio evidence \\the voice assistant has  started \\listening OR video or audio \\evidence that the voice assistant \\has started listening  in the \\absence of a cue\end{tabular}}
    & \textbf{Missed Trigger} &  Users say something to trigger the voice assistant, but it fails to respond. \\
     &  & \textbf{Spurious Trigger} & Users do not say something to trigger the voice assistant, but it activates anyways. \\
     &  & \textbf{Delayed Trigger} & Similar to system latency, the users say something  to trigger the voice assistant, but it replies too late to be useful. \\
\hline
      \multirow{10}{*}{\begin{tabular}[c]{@{}l@{}}\textbf{Perception}\end{tabular}} & \multirow{10}{*}{\begin{tabular}[c]{@{}l@{}}Visual or audio evidence that \\speech is being incorrectly \\captured by the system. For \\example, being cut off by the voice \\assistant, witnessing it continue to \\listen once the users' speech is \\complete, clearly mishearing a \\word, or evidence of background \\noise and cross-talk.\end{tabular}} 
      & \textbf{Noisy Channel} & User input is incorrectly captured due to background noise. \\
     &  & \textbf{Overcapture} & The voice assistant captures more input than intended by either beginning to capture input too early or ending too late, and acting on external data not relevant to the users' request. \\
     &  & \textbf{Truncation} & System does not fully capture users' speech, by either beginning to capture input too late or ending too early. \\
     &  & \textbf{Transcription} & System generates a transcription error, often in the form of similar sounding words. \\
\hline
    \multirow{9}{*}{\begin{tabular}[c]{@{}l@{}}\textbf{Understanding}\end{tabular}} & 
    \multirow{9}{*}{\begin{tabular}[c]{@{}l@{}}Suspecting that audio was \\correctly captured but not mapped \\to the correct action. For example, \\receiving a response indicating \\inability to complete an action that \\has worked in the past, or \\receiving a response that is \\plausible but not correct for the \\intention of the input.\end{tabular}}
    & \textbf{Ambiguity} & There may be several interpretations of the users' intent, and the system responds in a way that is plausibly accurate but not correct for the users' intent. \\
     &  & \textbf{Misunderstanding} & The system maps the users' input to an incorrect action, perhaps with some correct inference on the users' intent, but not fully accurate. \\
     &  & \textbf{No Understanding} & The system fails to map the user's input to any known action or response. \\
     & & & \\
\hline
    \multirow{5}{*}{\begin{tabular}[c]{@{}l@{}}\textbf{Response}\end{tabular}} & \multirow{5}{*}{\begin{tabular}[c]{@{}l@{}}Finally, assuming that the input \\was correctly captured and \\understood, was the response \\generated incorrect, unclear, not \\given, or otherwise wrong?\end{tabular}} & \textbf{Action Execution: No Action} & If the system listens to the full request, but then turns off before giving any type of answer or taking action. \\
     &  & \textbf{Action Execution: Incorrect Action} & The system gives information that is incorrect.\\
     & & & \\
\hline
\end{tabular}
\vspace{1em}
\caption{Qualitative codebook and description of the various failures that were collected. We checked each failure for failure type sequentially, starting by checking if it could be an attention failure and progressing through the types until we found one that fit. From failure type, we then assessed which failure source applied.}
\label{tab:codebook}
\end{table*}

\subsection{A Review of Existing HCI and NLP Datasets}

We first explored benchmark datasets in NLP, which contain a large number of either questions and answers~\cite{rajpurkar2016squad, rajpurkar2018know, reddy2019coqa}, or conversational dialogue~\cite{Gopalakrishnan2019, sun2018convdataset, zhang2018personalizing}. We found that existing NLP datasets do not cover the wide breadth of possible conversational failure cases due to their emphasis on correct data for training. Additionally, their focus on specific task performance, such as answering questions or dialogue generation, is more narrow than the variety of use cases for voice assistants. As training data relies on accurate task completion, these datasets did not contain failures. While testing these models produces a small percentage of errors (roughly 10\%), the types of failures could only fall in the response and understanding categories, as attention and perception failures are excluded from the context of training these types of models. This limited their usefulness for our purpose of understanding voice assistant failures that occur in use and their impact on user trust.

In addition to these benchmark datasets, we investigated datasets that incorporated spoken word speech patterns, such as the Spoken SQuAD dataset~\cite{lee2018spoken} and Disflu-QA dataset~\cite{gupta-etal-2021-disflqa}, as well as human-agent interaction datasets, such as the ACE dataset~\cite{Aneja2020ACEdataset}, the Niki and Julie corpus~\cite{artstein2018niki}, and a video dataset of voice assistant failures~\cite{cuadra2021look}. In these cases, we found that the datasets were still restricted to only failures at the understanding and response level~\cite{lee2018spoken, gupta-etal-2021-disflqa} or the context for the failures was very specific and did not necessarily capture the breadth of possible failures users experience~\cite{Aneja2020ACEdataset, artstein2018niki}. 
\citet{cuadra2021look}'s video dataset was the closest available fit for our needs, but we still found the use case of in-lab question-answering too narrow for our purposes.
Therefore, we decided to crowdsource a dataset of voice assistant failures from users, and use these failures when conducting our quantitative survey on user trust.

\subsection{Dataset Collection}
\subsubsection{Procedure}
Crowd workers were asked to submit three failures they had experienced with a voice assistant. They were asked about three specific types of failures out of a taxonomy of 12, which were randomly chosen and displayed in equal measure across all workers. The taxonomy of failures that we used to ask about specific types of failures was adapted from previous work by \citet{Hong2021nlpfailures}, and identifies failures due to attention, perception, understanding, and response, as shown in Table~\ref{tab:codebook}. Each question began by asking users if they could recall a time when their voice assistant had failed, based on the definitions in our taxonomy. For example, to capture missed trigger failures we asked ``Has there ever been a time when you intended to activate a voice assistant, but it did not respond?'' If so, we asked these workers to include 1. what they had said to the voice assistant, 2. how the voice assistant responded, 3. the context for the failure, including what happened in the environment, and 4. the frequency at which the failure occurred from 1 (rarely when I use it) to 5 (every time I use it). These were all presented as text entry boxes except for the frequency question, which was multiple choice. Crowd workers were additionally asked to optionally share an additional failure that they had not had the chance to share already. This was included to capture failures that did not fit any of the three the categories they were presented with, and we then categorized these failures according to our taxonomy.

Once we received these failures, we anonymized the type of voice assistant in the submitted examples, replacing activation words with ``Voice Assistant'' for consistency. We then edited grammatical and spelling errors for clarity. We also removed failures if they were not on-task, unclear, or exact repeats of other submitted failures. Finally, we noticed that some of the categories the users submitted the failures under were incorrect, so we re-categorized the failures according to the codebook we developed as outlined in Table~\ref{tab:codebook}.
Two raters iteratively coded 101 submitted failures, with a final coding session achieving an interrater agreement of 70\%. One researcher then went back and coded the entire dataset in its entirety. In total, our finalized dataset contains 199 failures across 12 categories, submitted by 107 unique crowd workers.

\begin{table*}[htbp]
\small
\centering
\begin{tabular}{p{0.14\textwidth}p{0.3\textwidth}p{0.22\textwidth}p{0.31\textwidth}}
\textbf{Failure Source} & \textbf{Context} & \textbf{What the User Said} & \textbf{How the Voice Assistant Reacted} \\
\hline
\hline
\textbf{Missed Trigger} &  I tell her to set a timer for ten minutes, I was alone and no one present at the moment. & Voice Assistant, set a timer for 10 minutes. & {[}No response.{]} \\
\textbf{Spurious Trigger} & I was having a conference call with my team, and I was calling my coworker Sherry. The voice assistant mistakenly got turned on. & {[}While talking to the coworker{]} "Can you share your screen?" & {[}Responding to the conversation with the co-worker.{]} "One moment, let me help you with that" \\
\textbf{Delayed Trigger}  & It happened while I was driving a car. & Voice Assistant, show me the route to the national park. &{[}The voice assistant takes so much time to respond that before it can respond, you once again ask the route.{]} \\
\hline
\textbf{Noisy Channel}  & My children were playing in the background and the dog was barking, and I had to raise my voice and try several times to be heard by my phone even though it was inches from my face. & Voice Assistant, what's the weather? &  {[}It didn't realize that my request had ended {]} and kept spinning.] ``I'm sorry, I didn't quite understand you.'' \\
\textbf{Overcapture} & I was telling it to turn off the lights. I was the only one there. & Voice Assistant, turn off the lights.& {[}It continues listening for so long that you turn them off yourself.{]} \\
\textbf{Truncation}  & I asked the voice assistant to calculate a math question, but it cut me off.& Voice Assistant, can you multiply 54, 39, 33, and 22? & "54 times 39 times 33 is 69,498." \\
\textbf{Transcription}  & I asked for the weather conditions in the city I live in. No others were present except for me. & Voice Assistant, what is the temperature in Murrieta, CA today? & "The temperature in Marietta, Georgia today is 65 degrees Fahrenheit." \\
 \hline
\textbf{Ambiguity}  & I was at home, alone, watching UFC and asked how old a fighter was. & Voice Assistant, how old is Johnny Walker? & "Johnny Walker was founded in 1865."  {[}It referred to the whiskey company instead of the fighter.{]}\\
\textbf{Misunderstanding} & I asked it to play the theme from Halloween. I was sitting with my mother. & Voice Assistant, play the theme song to the movie Halloween. & {[}It plays a scary sounds soundtrack instead of the song.{]} \\
\textbf{No Understanding} & I was trying to run a routine to wake up my kids. & Voice Assistant, wake up the twins.& "Sorry, I don't know that." {[}However, I've set up a routine for "wake up the twins" that has worked in the past.{]}\\
 \hline
\textbf{Action Execution: No Action} & I asked when a movie was coming out in theaters, and it kept spinning its light over and over. & Voice Assistant, when does Shang-Chi come out in theatres? & {[}Pauses for a really long time, then turns its lights off and does not respond.{]}\\
\textbf{Action Execution: Incorrect Action}  & I was at home, in my living room, alone. I was trying to find out how long Taco Bell was open.& Voice Assistant, when does the Taco Bell on Glenwood close. & "Taco Bell is open until 1am".  {[}Upon driving to Taco Bell, I realized it closed at 11:30pm.{]}
\end{tabular}
\caption{Table of voice assistant failures users submitted, including the context for the failure, what the user said, and what the voice assistant said.}
\label{tab:ex-fail}
\end{table*}

\subsubsection{Crowd Worker Characteristics}
We used Amazon Mechanical Turk to recruit the crowd workers.
In total, 107 crowd workers contributed to our dataset. We required workers to have the following qualifications: a HIT Approval Rate over 98\%, over 1000 HITs approved, AMT Masters, from the United States, over the age of 18, and voice assistant users on at least a weekly basis. The plurality of users were in the age range of 35-44 ($n=46$), followed by 25-34 ($n=32$), and 45-54 ($n=16$), with the rest falling in 55-64 ($n=8$), 18-24 ($n=1$), and 1 preferring not to answer. Fifty-eight crowd workers were men, 44 were women, 1 preferred not to answer, and 1 identified as both a man and a woman. They used commercial voice assistants such as Amazon Alexa ($n=59$), Google Assistant ($n=62$), and Apple's Siri ($n=40$), with many using some combination of the three ($n=47$). 91 crowd workers were native English speakers, and 13 were not. The plurality identified as White ($n=58$), and 39 identified as Asian. Three crowd workers did not provide any demographic information. The task took 15-20 minutes to complete on average, and they received \$5.00 USD compensation.

\subsubsection{Final Dataset}
In total, our finalized dataset contained 199 failures from 107 users across 12 different types of failures according to the taxonomy based on \citet{Hong2021nlpfailures}, as updated in Table~\ref{tab:codebook}.
The failures we received most often were due to misunderstanding ($n=38$), missed trigger ($n=25$), and noisy channel ($n=22$). Users least often submitted failures for truncation ($n=7$), overcapture ($n=7$), and delayed triggers ($n=8$). Most crowd workers submitted failures saying that they happened ``rarely when I use it'' ($n=87$) or ``sometimes when I use it'' ($n=84$). Example failures across the 12 categories can be found in Table~\ref{tab:ex-fail}.

On average, the highest frequency of failures occurred for no understanding ($m=2.15$, sometimes when I use it, $sd=0.67$) and action execution: incorrect ($m=2.00$, sometimes when I use it, $sd=0.88$). The rest of the failure sources had an average reported frequency between 1.0 (rarely when I use it) and 2.0 (sometimes when I use it). The lowest frequency failures were due to delayed triggers ($m=1.25$, $sd=0.46$) and ambiguity ($m=1.39$, $sd=0.78$).

We then used 60 of the failures from our dataset in our survey to quantify the impact of different failures on user trust. This is outlined in more detail in the following section.
This dataset has been open sourced\footnote{\url{ https://www.kaggle.com/datasets/googleai/voice-assistant-failures}} for researchers to use to answer future research questions related to voice assistant failures in the future.

\section{Interview and Survey Methods}
Once we had gathered and categorized our dataset of voice assistant failures, we were ready to answer our research question: how do voice assistant failures impact user trust? To do so, we first conducted exploratory interviews with 12 people to gather their thoughts, feelings, and behaviors after experiencing voice assistant failures. We used these findings and the failures collected in the dataset to then design and execute a survey. This quantified how various voice assistant failures impact users' trust, as measured by their perceptions of the voice assistant's ability, benevolence, integrity, and their willingness to use it for future tasks. Here, we describe the methods for both the interviews and survey, and we follow this by jointly presenting the results from both studies.
 
\subsection{Interview Methods}

\subsubsection{Interview Procedure}
Interviews began with questions about why the participants chose to start using voice assistants and what types of questions they frequently would ask of them. We asked for common times and places they would use their voice assistants to understand their general experience with voice assistants.

Once these were established, we asked participants to tell us about a time they were using their voice assistant and it made a mistake, in as much detail as they could recall. We asked what they had been trying to do and why, if others were present, and if anything else was happening in their environment. We probed for users' feelings once the failure occurred, and their perceptions about the voice assistant’s ability to understand them and give them accurate information. We asked participants what they did in the moment to respond to the failure. Finally, we asked questions about their use of the voice assistant in the aftermath, including how much they trusted it and if they changed any of their behaviors to mitigate future failures. All interviews were conducted remotely. 

\subsubsection{Interview Participants}
During recruitment, we asked participants to submit their demographic information, how frequently they used voice assistants and on what types of devices. We additionally required participants to write a short (1-3 sentence) summary of a time they encountered a failure while using their voice assistant. We selected participants based on demographic distribution and the level of detail they included regarding the failure.

All of our 12 participants lived in the United States. They used voice assistants at least 1-3 times a week ($n=2$), with the majority reporting using a voice assistant every day ($n=8$), and the rest ($n=2$) using it 4-6 times a week. The majority of participants used a voice assistant on their mobile device ($n=11$), and five of these participants also used a voice assistant smart home device. One participant only used a voice assistant smart home device. Participants reported using common commercial voice assistants such as Amazon Alexa ($n=2$), Google Assistant ($n=7$), and Apple's Siri ($n=8$).
Participants' ages ranged from 18 to 50, with the plurality ($n=5$) in the age range of 18-23. 3 of our participants were 41-50, 2 were 31-40, and 2 were 24-30. Six of our participants identified as women, five participants identified as men, and one participant identified as non-binary. Three participants identified as Asian, three identified as White, three identified as Black or African American, two identified as Hispanic, Latino, or Spanish origin, and one identified as both White and Black or African American. All of our participants spoke English as a native language. Participants were compensated with a \$50 gift card and each interview lasted roughly 30 minutes.

\subsubsection{Interview Analysis}
Interviews were transcribed in their entirety by an automated transcription service and analyzed via a deductive and inductive process~\cite{creswell2016qualitative}. We used deductive analysis to assess which types of failures these participants experienced. To ground our deductive analysis, we used the same codebook as we did for the dataset, as demonstrated in Table~\ref{tab:codebook}. We first identified instances in which participants were discussing distinct failures, and then applied our codebook to these instances. We used cues such as what was happening in their environment, and when appropriate, users' own perceptions of why the failure occurred. We began by first identifying if failures belonged in which of the four failure types: attention, perception, understanding, or response. First, to determine if there was an attention failure, we investigated if there was evidence that the voice assistant accurately responded to an activation phrase, as indicated by visual or auditory cues, or otherwise by the participant's narrative. Second, we evaluated if there was an error in perception, based on the participants’ assumption of if the voice assistant accurately parsed the input from the participant, our own assessment from their narrative, or other audio/visual cues. Next, assuming that the input was correctly parsed, we sought to understand if the voice assistant accurately understood the semantic meaning of the input (understanding failures), using the same process. Finally, assuming all else had been correctly understood, we assigned response failures, indicating that the voice assistant either did not take action or took the incorrect action in response to an accurately understood command. Once a failure type was determined, we then further specified the failure sources as noted in Table~\ref{tab:codebook}. We resolved disagreements both asynchronously and in meetings, through discussion and comparison, over the course of several weeks.

While conducting this analysis, we also inductively identified themes related to these failures' impact on future tasks and recovery strategies. To conduct this analysis, two researchers reviewed the twelve transcripts in their entirety, and one additional researcher reviewed five of these transcripts to further broaden and diversify themes. These researchers met over the course of several weeks to compare notes and themes, ultimately creating four different themes through inductive analysis. Of these themes, we report two due to their novelty, specifically as related to future task orientation and recovery strategies.

\subsection{Survey Methods}
To  quantify our findings from interviews, we developed a survey to explore users' trust in voice assistants following each of the twelve different types of failures from our taxonomy, as well as their willingness to use voice assistants for a variety of tasks in the aftermath.

\subsubsection{Procedure}
The survey contained a screener, the core task, and a demographic section. We required participants be over 18 years old, use their voice assistant in English, and use a voice assistant with some regularity to participate.
If participants passed the screener, they were required to review and agree to a digital consent form to continue.

The core task stated, ``\textit{The following questions will ask you what you think about the abilities of a voice assistant, given that the voice assistant has made a mistake. Imagine these mistakes have been made by a voice assistant you have used before. Please consider each scenario as independent of any that come before or follow it. This survey will take approximately 20 minutes.}'' Participants were then presented with 12 different failure scenarios, and they were asked to rate their trust in two separate questions.

The first question measured trust in voice assistants as a confidence score across three dimensions: ability, benevolence, and integrity. These were selected because prior work on trust has determined these elements explain a large portion of trustworthiness~\cite{ma2017self, mayer1995integrative}. In the context of voice assistants, ability refers to how capable the voice assistant is of accurately responding to users' input. Benevolence refers to how well-meaning the product is. And finally, integrity represents that it will adhere to ethical standards.

We asked participants to rate their confidence in voice assistants' ability, benevolence, and integrity, as a percentage on a scale of 0-100, with steps of 10, to replicate how prior work has conceptualized trust~\cite{ma2017self}. This was captured in response to the following statements:

\begin{itemize}[leftmargin=.3in]
    \item (Ability) This voice assistant is generally capable of accurately responding to commands.
    \item (Benevolence) This voice assistant is designed to satisfy the commands its users give.
    \item (Integrity) This voice assistant will not cause harm to its users.
\end{itemize}

The second question evaluated users' trust in the voice assistant to complete tasks that required high, medium, and low trust. To select these tasks, we ran a small survey on Mechanical Turk with 88 voice assistant users. We presented 12 different questions, which first gave an example voice assistant failure (one for each failure source), and then asked ``\textit{How much would you trust this voice assistant to do the following tasks}:'' give a weather forecast, play music, edit a shopping list, text a coworker, and send money. Users could choose that they would trust it completely, trust it somewhat, or not trust it at all.

There was not a significant difference in how much people trusted the voice assistant to play music compared to forecast the weather ($Z=2.06, p=0.078$). There was also not a significant difference in how much people trusted the voice assistant to edit a shopping cart or text a coworker ($Z=1.39, p=0.21$) as determined by pairwise comparisons, using $Z$-tests, corrected with Holm’s sequential Bonferroni procedure on an ANOVA of an ordinal mixed model. We found that there were significant differences between playing music, texting a coworker, and transferring money, with users having the most trust in the voice assistant playing music after a failure, less trust in texting a coworker, and still less in transferring money. Therefore, we selected playing music, texting a coworker, and transferring money to represent low, medium, and high levels of trust required. Therefore, after asking about ability, benevolence, and integrity, we asked participants how much they trusted their voice assistants to execute the following tasks: play music, text a coworker, and transfer money. These questions were displayed on a linear scale of 1 (``I do not trust it at all'') to 5 (``I completely trust it''), with steps of 1.

We completed the survey with an open-ended, optional question for participants to share anything else they would like to add. The survey concluded with demographic questions regarding gender, race, ethnicity, whether they were native English speakers, what type of voice assistants they used, and their general trust tendency as control variables.
General trust tendency was measured based on responses to the following: ``\textit{Generally speaking, would you say that most people can be trusted, or that you need to be very careful in dealing with people?}'' The options ranged from 1 (need to be very careful in dealing with people) to 5 (most people can be trusted). The questionnaire used for the survey has been submitted as supplementary materials.

\subsubsection{Materials from our Dataset}
To present each of the twelve failure sources in our survey, we drew from the dataset we had created. We selected five failures from each of the twelve categories. We required that these failures had been coded by two of the team members who were in agreement (see dataset examples in Table~\ref{tab:ex-fail}). We used random selection to determine which of the five possible failures was presented to each user for each failure source. These are denoted in the dataset ``Survey'' column.

\subsubsection{Participants} We recruited participants from Amazon Mechanical Turk. We first ran a small pilot ($n=27$) in which we determined that participants completed the survey in roughly 20 minutes on average, and we set the compensation rate at \$9 USD. After removing participants who did not pass the attention check or straight-lined, meaning they responded to every question with the same answer, we had a total of 268 participants. These participants were required to have the following qualifications: AMT Masters, with over 1000 HITs already approved, over 18 years old, live in the United States, an approval rate greater than 97\%, and they must not have participated in any of our prior studies. 
\begin{table*}[htbp]
\small
    \centering
    \resizebox{\textwidth}{!}{%
\begin{tabular}{l|rrrr|rrrr|rrrr}
& \multicolumn{4}{c}{\textbf{Ability}} & \multicolumn{4}{c}{\textbf{Benevolence}} & \multicolumn{4}{c}{\textbf{Integrity}} \\
\hline\hline
 & \multicolumn{1}{c}{$F$} & \multicolumn{1}{c}{df} & \multicolumn{1}{c}{residuals} & \multicolumn{1}{c}{$p$} & \multicolumn{1}{c}{$F$} & \multicolumn{1}{c}{df} & \multicolumn{1}{c}{residuals} & \multicolumn{1}{c}{$p$} & \multicolumn{1}{c}{$F$} & \multicolumn{1}{c}{df} & \multicolumn{1}{c}{residuals} & \multicolumn{1}{c}{$p$} \\
 \hline
(Intercept) & 9844.95 & 1 & 422.35 & \textless 0.001 & 8436.32 & 1 & 355.86 & \textless 0.001 & 10153.41 & 1 & 325.54 & \multicolumn{1}{l}{\textless 0.001} \\
General Trust & 3.07 & 4 & 286.67 & \multicolumn{1}{r}{0.017} & 1.78 & 4 & 299.1 & \multicolumn{1}{r}{0.133} & 4.69 & 4 & 311.17 & 0.001 \\
Failure Type & 17.17 & 3 & 2656.78 & \textless .001 & 8.87 & 3 & 2711.23 & \textless .001 & 20.56 & 3 & 2772.09 & \textless .001
\end{tabular}
}
    \caption{Voice assistant failure types significantly impacted users trust in voice assistants, across ability, benevolence, and integrity when controlling for their baseline trust tendencies, based on an ANOVA of three linear mixed models. Failure type was encoded as a categorical variable, and general trust was encoded as an ordinal value. Participant ID was a random, categorical variable.}
    \label{tab:confidence}
\end{table*}

The plurality of our participants were in the age range of 35-44 ($n=106$), followed by 25-34 ($n=68$), 45-54 ($n=52$), 55-64 ($n=33$), with 2-4 participants in each of the age brackets of 18-24, 65-74, and 75+. 134 of our participants identified as men, 132 identified as women, and 2 identified as non-binary genders. The majority of our participants were White ($n=210$), 21 participants were Black, and 15 were Asian. The rest of our participants identified as mixed race or preferred not to answer.

\section{Results: Trust in Voice Assistants after Failures}

In interviews, we found that participants reported failures across all four failure types and ten of the twelve failure sources. The only two failure sources that were not mentioned in interviews were missed triggers and delayed triggers in the attention failure type. To understand which types of failures most significantly impacted user trust, we analyzed how various failures impacted users' confidence in their voice assistant's ability, benevolence, and integrity. We used six mixed-linear regression models with log-normalized confidence in either ability, benevolence, or integrity as the numeric dependent variable. Note that there are two different levels at which we conduct the analysis. The first is at the four broad ``failure types'' level (attention, perception, understanding, and response). Then we drill down to the detailed 12 ``failure sources'' nested within each failure type.
Therefore, for each dimension of trust, we encoded failure type or failure source, as well as general trust tendency, as independent variables, so there were two regression models per dimension of trust. Failure type and failure source were encoded as categorical variables, and general trust tendency was encoded as an ordinal value. In all models, PID was encoded as a random, categorical variable.

\begin{figure}[b]
\centering
\includegraphics[width=\columnwidth]{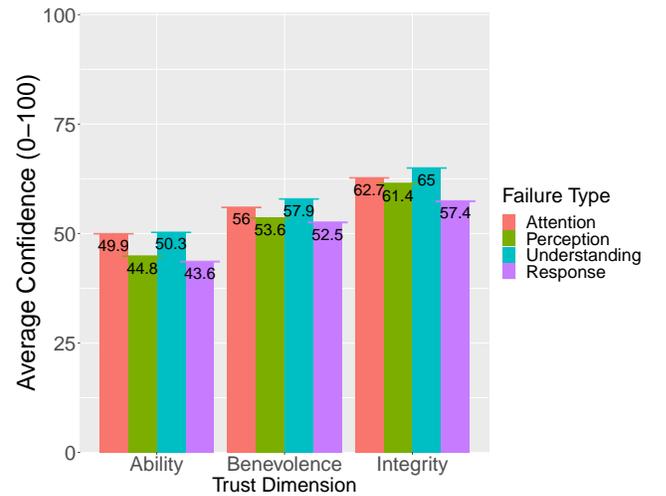}
  \caption{Average scores across the three dimensions of trust (ability, benevolence, and integrity) by failure type. Participants expressed higher confidence across three trust dimensions after encountering attention and understanding failures, compared to perception and response. Error bars display the confidence interval.}
  \label{fig:confidence-type}
  \Description{This figure contains a bar graphs, the y-axis is average confidence on a scale of 0 to 100. The x-axis is the trust dimensions of ability, benevolence, and integrity. The graph displays the confidence in each trust dimension by failure type: attention, perception, understanding, and response. We see that people are consistently more confident in their voice assistant after failures due to attention and understanding compared to perception and response.}
\end{figure}

\sloppy An ANOVA on the regression models revealed that failure type (attention, perception, understanding, response) does significantly impact perceptions of ability ($F(3, 2656.78)= 17.17, p<.001$), benevolence ($F(3, 2711.23)=8.87, p<0.001$), and integrity ($F(3, 2772.09)=20.56, p<.001$) when controlling for general trust tendency (see Fig.~\ref{fig:confidence-type} and Table~\ref{tab:confidence}). We found that the failure type ``Response'' (which includes action execution: inaction and action execution: incorrect action) more significantly deteriorated user trust in voice assistants across ability ($m = 43.6,\beta=-0.155$, $p < .001$), benevolence ($m=52.5, \beta=-0.072$, $p = 0.013$), and integrity ($m=57.3, \beta=-0.124$, $p < .001$), compared with failures due to ``Attention'' (which includes missed triggers, spurious triggers, and delayed triggers). Attention failures had a mean trust in ability of $49.9$, benevolence of $56.0$, and integrity of $67.7$ on the scale of $0$-$100$\%. We also found that failures due to perception significantly reduced users' confidence in voice assistant's ability ($m= 44.8, \beta = -0.122, p < .001$) and benevolence ($m = 53.6, \beta = -0.047, p = .05$), but had no measurable effect on integrity ($m = 61.4, \beta = -0.014, p = 0.484$) compared with attention failures. Failures due to understanding maintained higher user confidence in benevolence ($m=57.9, \beta = 0.054 , p = 0.031$) and integrity ($m=65.0, \beta = 0.063, p = 0.003$), but had no measurable effect on ability ($m=50.2, \beta = 0.015, p = 0.592$) compared with attention failures.

\begin{figure*}[htbp]
\centering
\includegraphics[width=\textwidth]{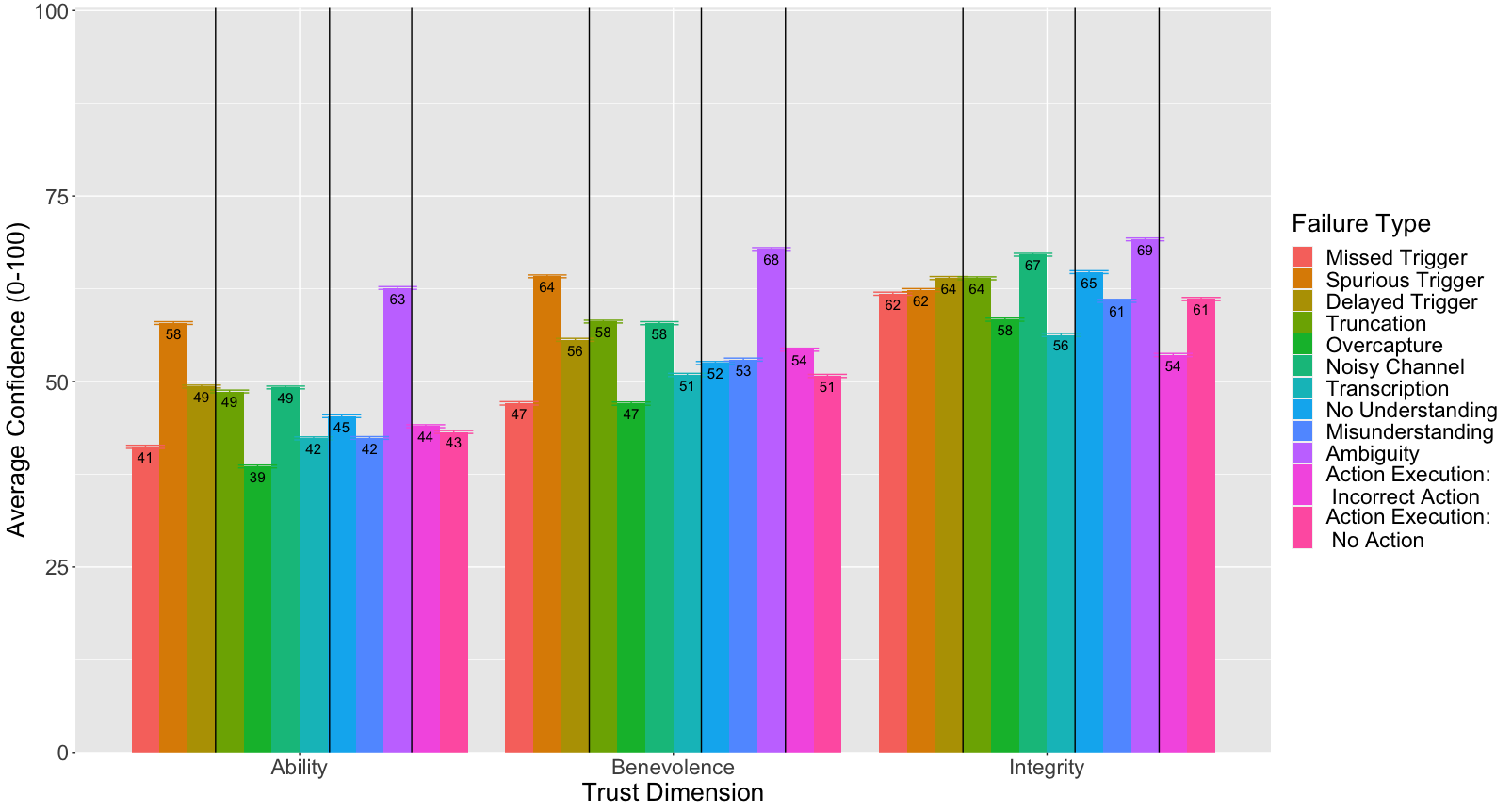}
  \caption{Average scores across the three dimensions of trust (ability, benevolence, and integrity) by failure type. Participants expressed higher confidence across three trust dimensions after encountering failures due to ambiguity and spurious triggers than they are of other failure types, especially missed triggers and overcapture failures.}
  \label{fig:confidence-source}
  \Description{This figure contains a bar graphs, the y-axis is average confidence on a scale of 0 to 100. The x-axis is the trust dimensions of ability, benevolence, and integrity. The graph displays the confidence in each trust dimension by the 12 failure sources. We see that people tend to be far more confident in the voice assistant after failures due to spurious triggers and ambiguity compared to other failure types. This is less apparent for integrity (which has a flatter shape, less variation).}
\end{figure*}%

Overall, response failures had the lowest average scores across ability, perception, and integrity. The most drastic difference between these categories is between failures due to understanding, which generally maintained the highest levels of trust in ability, benevolence, and integrity, as shown in Fig.~\ref{fig:confidence-type}. Therefore, in the analysis to follow that evaluates changes in trust across failure source, \textit{response: incorrect action} has been chosen as the reference variable, and all betas reported are in reference to this category. Below, we explore in more detail how users across both interviews and the survey responded to failures across the various failure sources.

\subsection{Attention Failures}
Attention failures are any failures in which a voice assistant does not accurately respond to an attempt for activation. These were the least commonly reported failures across interviews. In the survey, failures due to missed triggers were particularly harmful to users' confidence in voice assistants' ability ($m=41.2, \beta = -0.10, p=0.05$) and benevolence ($m=47.1, \beta =-0.228, p<0.001$). However, the impact on integrity was positive compared to the reference value (action execution: incorrect action) ($m=61.8,\beta = 0.194, p < 0.001$). None of the interview participants reported failures due to missed triggers. As all of the failures had a favorable impact on integrity compared to the reference value, we refrain from reporting it throughout the rest of the results. See Fig.~\ref{fig:confidence-source} and Table~\ref{tab:confidence-betas} for more details.

Only P7 and P12 reported experiencing attention failures in interviews, and they were both spurious triggers. As shown in Table~\ref{tab:codebook}, these are failures in which the voice assistant activates in the absence of an activation phrase. 
P7 reported that, 
\begin{table*}[htbp]
\small
\begin{tabular}{l|rrrr|rrrr|rrrr}
 & \multicolumn{4}{c}{\textbf{Ability}} & \multicolumn{4}{c}{\textbf{Benevolence}} & \multicolumn{4}{c}{\textbf{Integrity}} \\
 \hline \hline
 & \multicolumn{1}{l}{$betas$} & \multicolumn{1}{l}{se} & \multicolumn{1}{l}{$Z$} & \multicolumn{1}{l}{$p$} & \multicolumn{1}{l}{$betas$} & \multicolumn{1}{l}{se} & \multicolumn{1}{l}{$Z$} & \multicolumn{1}{l}{$p$} & \multicolumn{1}{l}{$betas$} & \multicolumn{1}{l}{se} & \multicolumn{1}{l}{$Z$} & \multicolumn{1}{l}{$p$} \\
 \hline
\textbf{(Intercept)} & 3.56 & 0.046 & 77.413 & \textless 0.001 & 3.796 & 0.048 & 79.083 & \textless 0.001 & 3.754 & 0.045 & 83.422 & \textless 0.001 \\
General Trust & 0.23 & 0.096 & 2.396 & 0.017 & 0.111 & 0.111 & 1 & 0.317 & 0.339 & 0.106 & 3.198 & 0.001 \\
\hline
Missed Trigger & -0.10 & 0.048 & -1.958 & 0.05 & -0.228 & 0.043 & -5.302 & \textless 0.001 & 0.194 & 0.037 & 5.243 & \textless 0.001 \\
Spurious Trigger & 0.36 & 0.046 & 7.761 & \textless 0.001 & 0.236 & 0.041 & 5.756 & \textless 0.001 & 0.212 & 0.037 & 5.73 & \textless 0.001 \\
Delayed Trigger & 0.16 & 0.046 & 3.391 & 0.001 & 0.043 & 0.042 & 1.024 & 0.306 & 0.249 & 0.036 & 6.917 & \textless 0.001 \\
\hline
Truncation & 0.16 & 0.046 & 3.522 & \textless 0.001 & 0.126 & 0.042 & 3 & 0.003 & 0.263 & 0.037 & 7.108 & \textless 0.001 \\
Overcapture & -0.13 & 0.047 & -2.809 & 0.005 & -0.193 & 0.042 & -4.595 & \textless 0.001 & 0.133 & 0.037 & 3.595 & \textless 0.001 \\
Noisy Channel & 0.14 & 0.046 & 2.978 & 0.003 & 0.076 & 0.042 & 1.81 & 0.07 & 0.326 & 0.036 & 9.056 & \textless 0.001 \\
Transcription & -0.06 & 0.047 & -1.213 & 0.225 & -0.101 & 0.042 & -2.405 & 0.016 & 0.093 & 0.037 & 2.514 & 0.012 \\
\hline
No Understanding & 0.06 & 0.047 & 1.17 & 0.242 & -0.043 & 0.042 & -1.024 & 0.306 & 0.28 & 0.037 & 7.568 & \textless 0.001 \\
Misunderstanding & -0.017 & 0.046 & -0.37 & 0.711 & -0.023 & 0.042 & -0.548 & 0.584 & 0.202 & 0.037 & 5.459 & \textless 0.001 \\
Ambiguity & 0.46 & 0.046 & 9.913 & \textless 0.001 & 0.302 & 0.042 & 7.19 & \textless 0.001 & 0.361 & 0.036 & 10.028 & \textless 0.001 \\
\hline
No Action & -0.003 & 0.047 & -0.064 & 0.949 & -0.09 & 0.042 & -2.143 & 0.032 & 0.186 & 0.037 & 5.027 & \textless 0.001 \\
\hline
\end{tabular}
\caption{The results of three mixed-linear regression models, demonstrating how voice assistant failures impact users' trust in voice assistants' across ability, benevolence, and integrity. Reference failure source: Incorrect Action.}
\label{tab:confidence-betas}
\end{table*}

\begin{quote} ``\textit{I feel like in conversation if I have it plugged in and there's like multiple people in the room, and they're talking or whatever, I think sometimes it may hear an [activation phrase] where it's not. And if that's happened, where it's activated like once or twice completely out of nowhere, and that hasn’t upset me or anything, but it's, it was just like, I didn't say [an activation phrase]. Why are you activating? What's happening? Why are you doing this?}'' \end{quote} 

P7 additionally said they were working and``\textit{It must have heard an [activation phrase] somewhere in there. And then it started speaking while I was trying to do my [job], and I had to like stop and be like, hey, stop.}'' They said, ``\textit{It would really piss [me off].}'' Similarly, P12 reported that these types of failures were ``\textit{irritating but funny at the same time}.'' They said they were funny ``\textit{because sometimes like, when you're usually calling [the voice assistant] she’ll take a longer time to respond, but when you're not talking to it, it automatically pops up\ldots Like, I’m not talking to you, but you could answer me when I'm talking to you.}'' As demonstrated in Fig.~\ref{fig:confidence-source} and Table~\ref{tab:confidence-betas}, failures due to spurious triggers had a more favorable relative impact on users' impressions of trust in the voice assistant's ability ($m=57.9, \beta = 0.36, p<0.001$) and benevolence ($m=64.2,\beta = 0.236, p<0.001$). Overall, it appears that these are one of the least detrimental failures to users' trust.

Similarly, failures due to delayed triggers were favorable to users perceptions of ability ($m= 49.3, \beta = 0.16, p = 0.001$) relative to the reference variable (response: incorrect action). Delayed trigger failures are defined as failures in which the voice assistant experiences latency when activating, to the point of potentially, but not necessarily, providing a correct response too late to be useful. They had no measurable effect on benevolence ($m= 55.6, \beta = 0.043, p = 0.306$). None of the participants reported a failure due to a delayed trigger in interviews.

\subsection{Perception Failures}
Users reported failures across all four failure types reported in Table~\ref{tab:codebook}, including truncation, overcapture, noisy channel, and transcription. Perception failures indicate that the voice assistant did not accurately capture the users' input. Transcription was by far the most common failure source, contrasted with only one failure recorded per truncation, overcapture, and noisy channel.

Truncation failures indicate that the voice assistant stopped listening to input too early, and only acted on some of the user's intended input. P12 reported that ``\textit{I use [a voice assistant] to send messages and stuff, and sometimes it would write the text for some of the words, but not all of the words. So it takes me longer than expected to send a message, because it will take a little bit of the words and not fully listen.}'' They said, ``\textit{it’s aggravating, very annoying}.'' Truncation failures had a favorable relative impact on perceptions of ability ($m=48.7, \beta =0.16, p<0.001$) and benevolence ($m = 58.1, \beta = 0.126, p = 0.003$). As shown in Fig.~\ref{fig:confidence-source}, these maintained higher relative trust compared to other failures in perception.

Overcapture failures indicate that the voice assistant has listened beyond the point that a user has given their input. As P8 said, sometimes, ``\textit{it doesn't know when to search for what I said and just keeps listening without taking action, even though it shows it is listening.}'' They tried to make sense of this failure, saying ``\textit{I find that on different devices, the reaction time for it [is different]}.'' They said that, ``\textit{This is wasting my time. Which is only logically two to three minutes,}'' but they said, ``\textit{if you keep messing with it, it makes it worse.}'' Failures due to overcapture were particularly harmful to users' confidence in voice assistants' ability ($m = 38.5,\beta = -0.10, p=.05$) and benevolence ($m = 47.0, \beta =-0.228, p<0.001$), with the overall lowest means compared to all other failure types.

There was one instance in which a user thought that the failure they experienced was because of noise in the background, indicative of noisy channel failures. P9 said, ``\textit{Sometimes\ldots I’ll try to use a feature where it tries to identify like a song\ldots and it just won't be able to pick it up, and it'll just give me a message, like `Sorry, I could not understand that.'}'' They said, ``\textit{I get that it was loud\ldots I would think that it would, it should be able to understand. So I feel like that is a little annoying.}'' However, they said the failure did not impact how they thought about the voice assistant’s accuracy or ability, saying that ``\textit{it’s pretty accurate for the most part, for other things.}'' Noisy channel failures were considered to more favorably impact user perceptions of ability ($m= 49.2, \beta = 0.14, p = 0.003$), with no measurable impact on benevolence ($m=57.9, \beta = 0.076, p = 0.07$). As shown in Fig.~\ref{fig:confidence-source}, they achieved similar levels of trust as failures due to truncation.

Nine of our participants mentioned failures relating to transcription of their input, indicating that they did not believe the voice assistant accurately captured what they had said. These failures varied from not understanding the name of a musical group (P7), incorrectly transcribing a text message (P2), incorrectly transcribing a sequence of numbers (P4), not understanding angry, slurred, or mumbled speech (P3, P5, P9), and not understanding accents (P8) or other languages (P6, P9). P7 said it caused a ``\textit{tiny little bit of frustration}'' when it did not understand the musician they were requesting. However, they ``\textit{don't really demerit [the voice assistant] for that in particular because it's so good at everything else that it does.}'' However, when it came to using the voice assistants in other languages such as Spanish or French, ``\textit{there has not been a successful time where it's been it's been able to play that different song in a different language}'' (P9). This led the participant to think, ``\textit{that it -- it just has no ability to understand me in a different language}'' (P9). Failures due to transcription did not have a measurable impact on perceptions of ability in the survey ($m = 42.3, \beta = -0.06, p = 0.225$) relative to the reference variable, however they impacted trust more so than other failure sources within perception as shown in Fig.~\ref{fig:confidence-source}. Transcription failures did negatively impact perceptions of benevolence  ($m = 51.0, \beta = -0.101, p = 0.016$).

\subsection{Understanding Failures}
We found that participants submitted failures across all categories of understanding failures, as described below.

Failures due to no understanding resulted in a complete inability to map the input to an action or response. P6 said, ``\textit{I was trying to plan a vacation\ldots It was my friend's bachelorette party\ldots And I was like, [Voice Assistant], where's Lake Havasu? How far is it?\ldots And she's like, `Sorry. I didn't understand what you're saying.'}'' This led P6 to question, ``\textit{Why do I even use you?}'' However, they said that, ``\textit{for timers, it works really well}.'' No understanding failures did not significantly impact trust relative to the reference variable, in terms of ability ($m= 45.3, \beta = 0.06, p = 0.24$) or benevolence ($m= 52.5, \beta = -0.043, p = 0.306$).

Misunderstanding failures occurred when the voice assistant mapped the user’s input to an action that was partially, but not fully, accurate to their intent. For example, P4 explained that when they ask their voice assistant ``\textit{to `Take me home.' It usually directs me to my home, but on occasion, it shows me search results for the phrase `Take me home.'}'' Similarly, P1 explained how when using a voice assistant for online shopping, sometimes it would ``\textit{pull up the wrong item or, like, the wrong location}.'' They said they felt ``\textit{disappointed and frustrated.}'' Misunderstanding failures did not measurably impact perceptions of ability ($m= 42.4, \beta = -0.017, p=0.711$) or benevolence ($m= 52.9, \beta = -0.023, p=0.584$) relative to the reference variable.

Failures due to ambiguity were situations in which one could see several reasonable interpretations of one’s intent from the captured input, but the system failed to navigate the ambiguity. For example, P10 said, ``\textit{I was trying to get to Pizza Hut and\ldots it kept on telling me one in the nearby city instead of the one that's I believe like 10 minutes away from me. So I asked a couple of times, and then it didn't work, and that's when I just pulled out my phone and then just looked it up myself and left.}'' They said that they were ``\textit{a bit baffled, since normally, like when I ask [a voice assistant] for something, I get the response I would expect.}'' As demonstrated in Fig.~\ref{fig:confidence-source} and Table~\ref{tab:confidence-betas}, failures due to ambiguity were more favorable to users' impressions of the voice assistant's ability ($m = 62.6, \beta = 0.46, p<0.001$) and benevolence ($m = 67.9, \beta = 0.302, p<0.001$). Overall, these failures maintained the highest level of user trust.

\subsection{Response Failures}
There were two possible types of response failures. These included incorrect action, in which the system gives information that is incorrect, or no action, in which a voice assistant fails to respond at all. 

Incorrect action failures were times when the command seemed to be accurately understood, but the information provided in response was incorrect. For example, P1 said that sometimes they would use ``\textit{the voice assistant to give me the best route to get to the location.}'' While it would usually accurately respond to this command, sometimes, ``\textit{it will give me a really like roundabout way, like really time-consuming way.}'' As shown in Fig.~\ref{fig:confidence-source}, failures due to incorrect action resulted in a relatively average perception of ability ($m = 44.0$) and benevolence ($m = 54.3$), and the lowest perception of integrity ($m=53.6$).

Multiple users experienced failures due to no action, in which the voice assistant completely fails to respond to the input. P2 said, ``\textit{I did have a couple times that was also frustrating\ldots I would say `Reply' [to a text message]. And I would talk and nothing would get sent. And like, my hands are literally covered in stuff because I'm rolling these cookies out, and I had to stop what I'm doing, go back to my phone, and actually like manually text.}'' Another participant experienced failures due to no action, saying that ``\textit{This morning where I woke up. I  said, [Voice Assistant], what's the weather outside? And it loaded for the first few seconds\ldots and then after a couple of seconds, it said, `There was an error. Please try again in a few minutes.' I wait one or two seconds, then I'll ask it again, and it gives me the information}'' (P10). This participant said that because the information has been ``\textit{accurate},'' they ``\textit{would still trust it to a very high degree.}'' Failures due to no action had no measurable relative impact on ability ($m = 43.2, \beta = -0.003, p = 0.949$) and had a slight but significant negative impact on benevolence ($m = 50.7, \beta = -0.09, p = 0.032$) compared to incorrect action.

\section{Responses to Failures and Future Use of Voice Assistants}

Users described a variety of strategies for mitigating failures, given that they did occur. In some cases, users described completely stopping their use of a voice assistant for a particular task. For example, after encountering a truncation failure while using the voice assistant to send a text message, P12 said that they either ``\textit{have to redo it, or I just, like, don't do it at all}.'' Eventually, P12 said that they stopped encountering that failure because they ``\textit{barely use it}'' for that same task anymore. So while some users felt like they ``\textit{don't sweat it too much}'' (P5) when a voice assistant failed at a task, others felt like they would use it ``\textit{not as much}'' (P2) for those same tasks. 

We found that the pattern of continuing to use a voice assistant in general but excluding the tasks that resulted in a failure, at least for a short period of time, was consistent across many different types of failures, including transcription, misunderstanding, and ambiguity. For example, P2 said that they needed to be careful using a voice assistant, because sometimes they would say a name and ``\textit{it would come up [with] a different name.}'' They said that following an incident like that, 

\begin{table*}[htbp]
\small
\begin{tabular}{l|rrrl|rrrl|rrrl}
 & \multicolumn{4}{c}{\textbf{Playing a Song}} & \multicolumn{4}{c}{\textbf{Texting a Coworker}} & \multicolumn{4}{c}{\textbf{Transferring Money}} \\
 \hline
 & \multicolumn{1}{c}{$\beta$} & \multicolumn{1}{c}{$se$} & \multicolumn{1}{c}{$Z$} & \multicolumn{1}{c}{$p$} & \multicolumn{1}{c}{$\beta$} & \multicolumn{1}{c}{$se$} & \multicolumn{1}{c}{$Z$} & \multicolumn{1}{c}{$p$} & \multicolumn{1}{c}{$\beta$} & \multicolumn{1}{c}{$se$} & \multicolumn{1}{c}{$Z$} & \multicolumn{1}{c}{$p$} \\
 \hline
Ability & 0.048 & 0.003 & 16.000 & \textless 0.001 & 0.064 & 0.003 & 21.333 & \textless 0.001 & 0.076 & 0.005 & 15.2 & \textless 0.001 \\
Benevolence & 0.043 & 0.003 & 14.333 & \textless 0.001 & 0.028 & 0.003 & 9.333 & \textless 0.001 & 0.017 & 0.005 & 3.4 & 0.001 \\
Integrity & 0.019 & 0.002 & 9.500 & \textless 0.001 & 0.024 & 0.003 & 8 & \textless 0.001 & 0.03 & 0.004 & 7.5 & \textless 0.001 \\
General Trust & -0.146 & 0.108 & -1.352 & 0.176 & -0.094 & 0.134 & -0.701 & 0.483 & -0.192 & 0.218 & -0.881 & 0.378
\end{tabular}
\caption{The results of three mixed-ordinal regressions modeling user trust in the voice assistant to execute the task based on their perceptions of the voice assistant's ability, benevolence, and integrity. We did not include cut point calculations and state 1 calculations in the table for ease of interpretability.}
\label{tab:future-tasks}
\end{table*}

\begin{quote}``\textit{I would still use [the voice assistant]. I think what would happen though is like you kind of build up that trust\ldots So the next couple times I would go into my contacts and hit the button myself, you know, and then like if I was walking to my car and get my keys in one hand, and it's been a while. So, you know, let me try this again. Like I think that's something where you kind of have to like, build the trust back up and give it another try. At least that's what I do.}'' \end{quote}

\noindent P12 echoed this, saying, ``\textit{Let's say you're opening Spotify or something like that. I think it will probably go on command, rather than sending a message\ldots different tasks, you know, it has a different trust level.}'' P5 had a similar sentiment, saying ``\textit{I think the problem with the most voice assistant is, if I tried to give it a complex search query, it doesn't really understand me, or it gets frustrating and I just I'm going to go ahead and type in whatever it is I'm looking for.}''  Even when failures were mitigated in the moment, users remained wary of using their voice assistants for the same tasks. 

Interestingly, sometimes users would continue to use their voice assistant for the same general task following a failure, but they would make slight changes to their use. For example, P1 encountered a misunderstanding failure while trying to shop for a sweater online, and they started to ``\textit{rely on it a little less, and do more searching on my own.}'' They said that ``\textit{for future reference, I would just remember to not use it to do certain tasks and do certain tasks on my own, [especially] when I look for an item that's difficult to find}.'' However, in the meantime, ``\textit{I would just ask for other tasks.}'' For P1, this included ``\textit{looking for other items other than this sweater. I would tell her to search for like grocery items and do some comparison shopping online.}'' Shopping for different items was distinct enough to maintain this user's trust. P7 experienced a similar situation, in which they encountered a transcription error, which they mitigated by spelling the name of ``\textit{hyerpop duo 100 gecs}'' as ``\textit{G-E-C-S}.'' They said this correct helped so that ``\textit{[the assistant did] understand what I was saying.}'' Even though they had experienced a failure for that particular artist, they ``\textit{continue to do that [use it to play songs] to this day. It's a very good music player,}'' but they are ``\textit{a little weary when it comes to certain musicians that I feel that\ldots [the voice assistant] would have trouble understanding.}''

Users often made sense of the failures based on the perceived task complexity. P12 thought that the task that they had the highest trust in was ``\textit{to open like apps},'' followed by ``\textit{calling somewhere}.'' They explained that, ``\textit{I want to put that as number one, but sometimes, like the way the contact name is, is not registered. Like, you know the way for you to say it, it's not how like the voice [assistant] says it.}'' P2 similarly evaluated the voice assistant, saying, ``\textit{the best thing is picking up website information}.'' However, they similarly said ``\textit{to get more personalized messages, contacts, and that sort of thing, you have to be really careful what you say and how you say it.}''

Because of these findings, we hypothesized that users' trust in voice assistants after failures would affect their willingness to use it for different tasks to differing degrees.
As shown in
Table~\ref{tab:future-tasks}, 
we used three mixed-ordinal regressions to model trust in these three tasks, with scores for confidence in the voice assistant's ability, benevolence, and integrity as the independent variables. Trust in the voice assistant to play a song, text a coworker, and tranfer money was encoded as an ordinal value. Confidence in the voice assistant's ability, benevolence, and integrity were encoded as numerical values. General trust tendency was encoded as a numerical value and PID was encoded as a random categorical value. We found that user perceptions of voice assistant ability, benevolence, and integrity positively correlated with their willingness to use the voice assistant for future tasks. Overall, people were moderately trusting of their voice assistant to play a song ($m=3.29, sd = 1.30$), less trusting of their voice assistant to text a coworker ($m=2.34, sd=1.18$), and least trusting of their voice assistant to transfer money ($m=1.56,sd=0.91$).

In particular, perceptions of ability had a stronger effect on people's willingness to use the voice assistant to play a song ($\beta = 0.048, p < .001$) compared with benevolence, which also significantly impacted willingness to use the voice assistant to play songs, but to a slightly lesser degree ($\beta = 0.043, p < .001$). Integrity was even less influential, though still significantly positively correlated without how much people trusted their voice assistant to play a song ($\beta = 0.019, p < .001$). This pattern was repeated for texting a coworker and transferring money as well, with ability being most strongly positively correlated with people's willingness to trust the voice assistant to execute these tasks, followed by benevolence, and then integrity.

\section{Discussion}

With interviews, a survey, and a crowdsourced voice assistant failures dataset, we conducted a mixed-method study of voice assistant failures and how they impact user trust in and future intended use of voice assistants.
As the underlying technology for voice assistants continues to improve in accuracy and ability, and its applications become increasingly high stakes to human health and well-being~\cite{sezgin2020readiness, mehandru2022reliable, yang2021clinical, de2020reducing}, we discuss our findings with the goal of improving user trust and long-term engagement in voice assistants.

Our users consistently relied on their voice assistants to find information and execute tasks across varying levels of complexity. Similar to prior work~\cite{luger2016like}, those who wanted to use a voice assistant consistently for tasks which might result in failures have developed complex mental models of which tasks they can trust their voice assistants with. Unlike prior work~\cite{luger2016like}, people often did not necessarily entirely abandon the use of their voice assistant after it failed at complex tasks, even after repeated failures. Many users considered the accuracy of their voice assistants so consistently high that they could forgive failures and continue engaging with those tasks after a short period of time. While trust in the complex tasks was being repaired, many participants continued using their voice assistants for tasks they considered more simple, such as information retrieval and playing music. 

We find that failures that lead users to feel like they have wasted time, such as those due to missed triggers and overcapture, tend to lead to more deteriorated perceptions of ability and benevolence. This is contrasted with scenarios in which users have more understanding of \textit{why} the voice assistant failed, such as those due to ambiguity and transcription, which users generally felt like they could work around or anticipate. However, if the failure due to transcription was believed to be due to using the device in another language, this caused abandonment of the voice assistant in that language. Similarly, users did not feel like they lost out on the advantages of using voice assistants when spurious trigger failures occurred, so they were less damaging to perceptions of ability. The single most damaging failure source to voice assistant integrity was action execution: incorrect action, as participants were more skeptical of the claim that the voice assistant would not cause harm following these failures.

Prior work has pointed to ways that trust can be repaired when failures do occur. \citet{cuadra2021my} showed that when a voice assistant proactively attempts to acknowledge a failure and repair trust, this increased people's perception of its intelligence. Additionally, \citet{mahmood2022owning} has found that failure mitigation strategies such as apologies were effective in restoring perceptions of likability and intelligence of a voice assistant after a failure. \citet{Xiao2021spotify} demonstrated that situating the voice assistant as a learner, and helping users understand when to give feedback to the voice assistant, improved users' perceptions of the voice assistant. \citet{Fischer2019progressivity} encourages voice assistant responses to support progressivity of the conversation, especially when the response does not help the user. Our work shows that users naturally repair trust with their voice assistants by relying on it for different tasks following a failure, or the same task but on a different topic, such as online shopping for different items or playing music by other artists than those that caused a failure.

Quantitatively, we established that certain types of failures are more critical than others. This insight can be used to help prioritize the failure recovery strategies across HCI and NLP that are most effective for regaining trust.
For example, self-repair for voice assistants such as \citet{cuadra2021my} employed may be most useful in situations where the voice assistant has failed because of a missed trigger or overcapturing users' input.
In addition, we can also try to identify the specific components in the voice assistant technology stack that cause critical failures, and leverage techniques in NLP robustness to improve how these models perform during user interactions.
For example, noisy channel and transcription failures can be modeled as small perturbations to the input, which is well researched \cite{ebrahimi-etal-2018-hotflip,belinkov2018synthetic}.
Reliable transcription may also be important to address by speech recognition modules, especially for low resource languages~\cite{magueresse2020low}.

Our open-sourced dataset has also provided concrete and comprehensive example failures (199 real-world sourced examples with context, query, and response) for future researcher to reuse to develop failure mitigation strategies, along with a refined taxonomy for classifying voice assistant failures, as supported by prior work. While prior work~\cite{Hong2021nlpfailures} was useful in helping NLP practitioners anticipate and plan for failures across many types of NLP technologies, our dataset specifically addresses failures that occur with voice assistants. We anticipate this will allow future researchers to use human-centered example failures when conducting research related to voice assistant failures, trust, and mitigation strategies.

\section*{Limitations}
There are a few methodological limitations of our study, which we detail here. First, our dataset collection and interviews relied on retrospectives and recalling failures, rather than observing them \textit{in situ}. This subjects our data to recall bias, and our results should be interpreted in this light. For instance, none of the interview participants recalled failures due to missed triggers or delayed triggers in interviews, although missed triggers were considered relatively damaging to perceptions of ability and benevolence in the survey. Additionally, our survey relied on collecting users' feedback regarding hypothetical scenarios. Future work may build on our findings by using our dataset to systematically introduce failures and capture the resulting impact on user trust via ESM or diary study. Our sample of participants were also frequent voice assistant users, which indicates that they likely forgave errors more easily than other populations~\cite{luger2016like}.
Additionally, we did not address the use of conversational agents through interfaces other than voice, such as embodied conversational agents or text-based conversational agents. As embodied and text interfaces have more potential affordances with which users can judge and interact with the system~\cite{bickmore2001relational, bickmore2005social}, the impact of failures may not perfectly generalize to these use cases. 

\section{Conclusion}
In conclusion, through a mixed-method study, we found that voice assistant users experience a multitude of failures, ranging from a voice assistant incorrectly triggering to responding in a way that does not address users' needs.
These different types of failures do differentially impact users' trust, which in turn affects intention to use their voice assistants for tasks in the future.
In particular, we find that failures due to spurious triggers and ambiguity are less detrimental to user trust than failures due to incorrect action execution, missed triggers, or overcapture.
We additionally find that people rebuild their trust in voice assistants through simple tasks, such as playing a song, before resuming using their full voice assistant functionality after a failure has occurred.
We also contribute a dataset of 199 failures, to help future researchers and practitioners build on our work.
By further working to understand, prevent, and repair voice assistant failures, we hope to build voice assistant users' trust in these devices and allow them to benefit from the increasing and varied functionality they provide.

\newpage
\bibliographystyle{ACM-Reference-Format}
\bibliography{ref.bib}

\end{document}